# Development of On Board Computer for a Nanosatellite

**Saurabh M. Raje[a]*, Abhishek Goel[b], Shubham Sharma[c], Kushagra Aggarwal[d], Dhananjay Mantri[e], Tanuj Kumar[f]**

[a, e] *Department of Computer Science and Information Systems, BITS Pilani, Pilani 333031, India*
[b, c, d, f] *Department of Electrical and Electronics Engineering, BITS Pilani, Pilani 333031, India*
* Corresponding Author

**Abstract**

A group of undergraduate students from BITS-Pilani are building a nanosatellite whose objective is to perform hyperspectral imaging of the oceans. This has never been carried out by a nanosatellite. The spectral distribution shall help in categorizing various phytoplankton in the oceans. This is used to study the carbon cycle of the oceans, which has an impact on the marine life. This paper describes the process of conceptualization, design and testing of the Onboard Computer (OBC) Software for a 3U nanosatellite. The on-board computer of the satellite is responsible for controlling the activities of all other subsystems, initiating dataflow between onboard hardware and performing mission critical computations like image compression. Control algorithms for fine pointing, sun pointing, ground pointing for payload operation and idle state detumbling run on the OBC. The actuation associated with this is carried out by interfacing magnetorquers and reaction wheels with the OBC. The software of the onboard computer is implemented on a Linux based operating system run on the ARM Cortex A9 processor which is part of the Zynq-7000 SoC. A Field Programmable Gate Array (FPGA) is used specifically for image compression. The compressed image is stored in a serial flash memory shared between the camera and the FPGA. The architecture comprises of a system-wide I2C bus to which various sensors like magnetometer, temperature sensor, Inertial Measurement Unit (IMU) etc. are interfaced. The collected data is used for logging followed by downlink, and as input to algorithms used for pointing and detumbling. An SPI interface is used between the Power Subsystem microcontroller and the On-Board Computer since a large amount of housekeeping data will have to be exchanged at high rates. Also actuators namely, reaction wheels and magnetorquers are actuated by current driver circuits which get the control signals from the OBC. The satellite is modelled as a Finite State Machine for software development. The states broadly fall under two categories, Normal and Emergency. Each state has a predetermined set of logical tasks to be run, which are abstracted as separate processes in the memory. State transitions take place by polling the health metrics of the satellite. However, hardware interrupts are implemented on selected peripherals which ensure a asynchronous switching to Emergency States for safety. A review of some common fault detection, isolation and removal methods used shall conclude the paper.

## 1. Introduction

While arriving at the hardware and software architecture for this paper, various successful on board computer (OBC) designs were reviewed. An alternate name for the OBC is command and data handling system (CDHS). This name better describes the purpose of this system. However, depending on the mission objectives, various satellites have also used additional computing units for their payload. Hence each OBC design has been reviewed keeping in mind the nature of the mission. A similar review conducted by the team at CalPoly has also been referred to [1]. Section 2 of this paper describes the hardware layout of the OBC, section 3 elucidates the various operational modes of the satellite, section 4 describes the software architecture and the levels of abstraction therein, section 5 describes the kernel level modifications and custom modules developed, and section 6 highlights some measures taken to inculcate fault tolerance in the system.

## 2. Hardware Architecture

The payload for this mission is a hyperspectral camera, which will capture a region in the Bay of Bengal. The number of bands of wavelengths (150 in this system) captured by camera decides the size of the image. The image size ranges around 400 megabytes and transmitting such a large amount of data from a nanosatellite is very difficult, considering the bandwidth constraints and flyby time over the ground station. This necessitates image compression on board. High performance compression coupled with low power consumption, and a soft real time constraint on the data handling and command execution hints at the use of multiple computing units which will be used for specific tasks and can be switched off independent of each other.

The OBC hence consists of a board for CDH, an FPGA for image processing pertaining to the payload, and the interfaces from the CDH board to the various sensors and actuators for exchanging data as per the corresponding subsystem requirements. The complete hardware block diagram is shown in Fig. 1.

*2.1 Computing unit for CDH*

The computing unit for CDHS is designed keeping in mind the use cases and the computational capacity of the other subsystems. Apart from data handling and task scheduling, this board also runs the control algorithms written to ensure the rotational stability of the satellite in orbit. Hence the actuators associated with the same (viz. reaction wheels and magnetorquers) are interfaced directly with this board. Since the telemetry subsystem has its own microcontroller, the transceiver is not interfaced with this board. Furthermore, the power subsystem also has its own microcontroller (MSP-430, developed by Texas Instruments). This will be responsible for booting up the OBC system.

The Zynq-7000 SoC developed by Xilinx has been chosen for the CDH unit. It consists of a dual core ARM Cortex A9 processor clocked at 800Mhz.

*2.2 Bus interfaces for CDH*

The CDH is interfaced to various sensors on the inter-integrated circuit (I2C) bus, owing to the ease of development and the availability of CMOS sensors with I2C interface for data read. SPI bus has been used in cases where relatively larger data is transferred. Specifically, in the interface between the CDH unit and the electrical and power subsystem microcontroller (MSP-430). This interface will provide the CDH with all the housekeeping data required for logging. SPI will also be used to transfer images from the flash memory to the FPGA for compression and then back to another flash memory.

*2.3 Computing unit for image processing*

The Programmable Logic (inbuilt to Zynq-7000) is used for image compression. This FPGA is programmed to read image data in bursts using a custom SPI soft controller from a flash memory, perform hyperspectral image compression using CCSDS123 lossless hyperspectral compression algorithm, and finally write the compressed data back to another flash memory [2]. This has been done in order to speed up the compression process by pipelining it at a hardware level. Furthermore, it is necessary to keep the CDH unit operational at all times for attitude stabilization and logging. Such a separate computing unit for payload ensures that the CDH unit can carry out its tasks asynchronously. Use of FPGA also necessitates radiation hardening measures. A literature review has been carried out and measures like scrubbing will be implemented [3]. Section 6 deals with this in more detail.

**3. Modes of operation**

The functioning of the satellite can be represented as a finite state machine in which there are a set of states/modes and depending on certain conditions or health parameters, (such as power level, angular frequency of rotation etc) the satellite will switch among the modes. There is a certain (disjoint) set of modes that represent anomalies ("Emergency Modes"). Polling the metrics mentioned above will determine which mode to switch to. However, in order to reduce the response time, switching to the emergency modes will be interrupt based. These interrupts will be generated by the relevant hardware or by relevant software which detect the crossing of critical thresholds. The interrupts may be handled by custom built drivers. Each of these modes has a set of tasks that need to be run when the satellite is in the given mode.

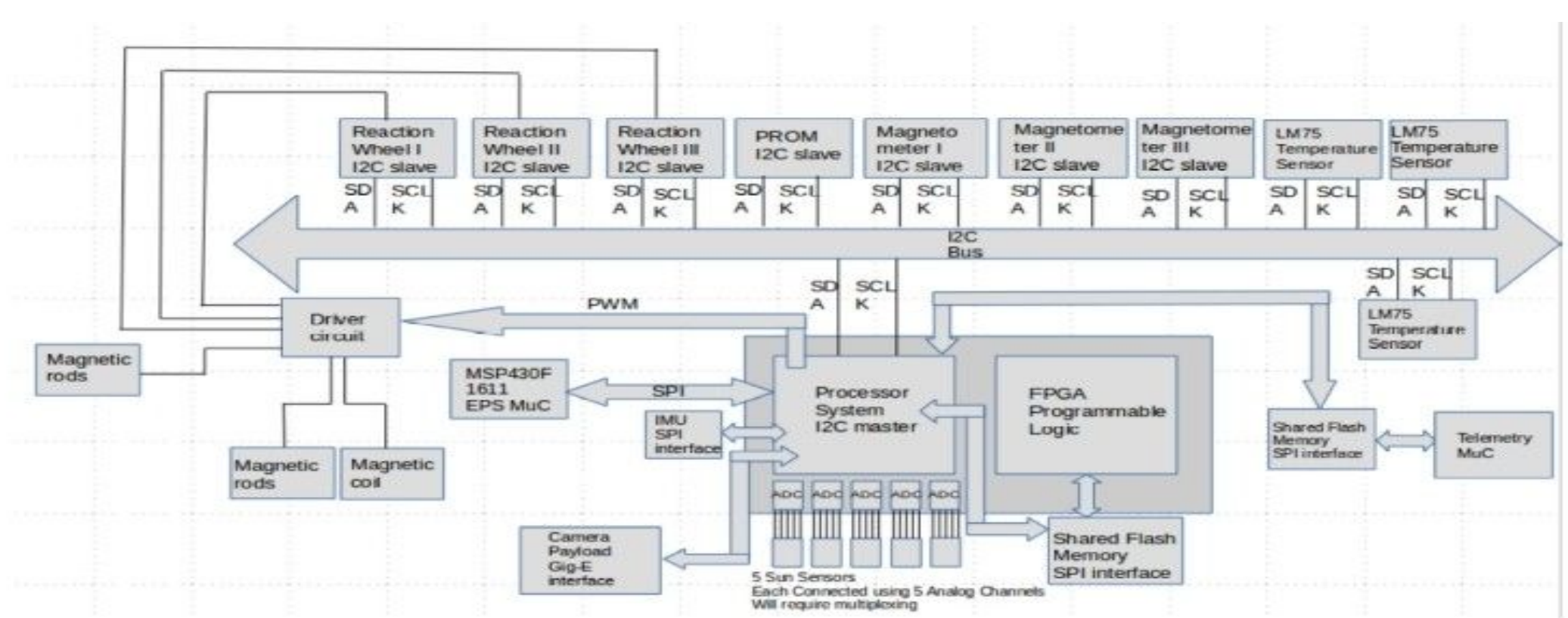

*Figure 1: Hardware block diagram*

## 4. Software architecture

The software is organized into 3 levels of abstraction above the Petalinux operating system used. In addition to this, some drivers have also been developed and integrated with the vanilla kernel. This architecture is in contrast with the multithreaded model used in AAUSAT-3 [4]. The following sections elucidate the various layers of abstraction in a top-down order and Fig 2. illustrates the same.

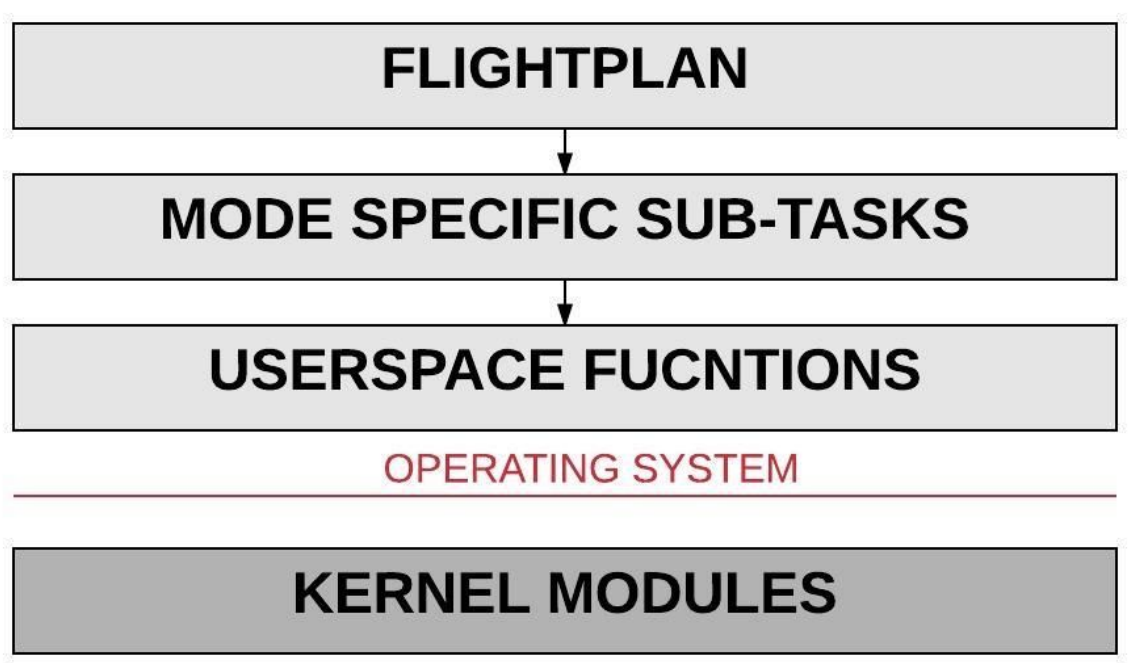

*Figure 2 : Software Abstraction Layers*

### *4.1 The flightplan process*

The topmost level of abstraction is a single process (called "Flightplan") that represents the mode of operation of the satellite, which in turn determines the tasks to be executed. The Flightplan process is responsible for scheduling, executing and tracking all the logical tasks that need to be done in a particular mode. This process is polymorphic in the sense that for every mode of operation, this process will superimpose different executable process images (using execv command) but will represent the same process in the memory. The Flightplan process forks and executes multiple child processes that form the middle layer described in the next subsection. This is done at a predetermined frequency of execution for each process/task. The tasks are put in a list, ordered by next time of execution. A task is forked and executed as a child process when it is at the front of the list and:

1. The current system time is greater than or equal to the time of next execution.
2. There is no child process running in the system that is doing the same task.

Once the process has been forked and executed, it is put back in the list as per the next time of execution. The list mentioned above also contains metadata about each child process including but not limited to its PID & frequency of execution. This is particularly useful when the process terminates. To handle the zombie process, a custom SIGCHILD handler has been implemented (non-blocking). This updates the metadata about the process by waiting on it. This concept of signal handling will be extended to use the SIGRTXXX signals to model software interrupts among the different processes [5]. It will successfully replace the need for any form of interprocess communication mechanism. Switching amongst the modes takes place by adding an extra node ("check" node) to the list mentioned above. When this node is at the front of the list, a set of parameters will be checked and conditional switching will take place among the modes. If the Flightplan decides to switch, it will have to wait for the existing children to terminate. This is another reason why the metadata of each child must be stored in the parent.

### *4.2 Mode-specific subtasks*

Various logically independent tasks need to be carried out in each mode as shown in Fig. 3. These tasks are implemented as separate processes in the memory. These tasks call a set of userspace functions that wrap around the driver attribute functions using the ioctl() call. This ensures that the developers can continue to develop them independently of the driver functions being implemented. It also makes debugging easier. The control algorithms that detumble the satellite are also implemented at this level. In a mode, only one control algorithm is run. As mentioned above, it uses the userspace functions to gather data from the sensors and also to send the Pulse Width Modulated (PWM) wave to a driver circuit for actuation.

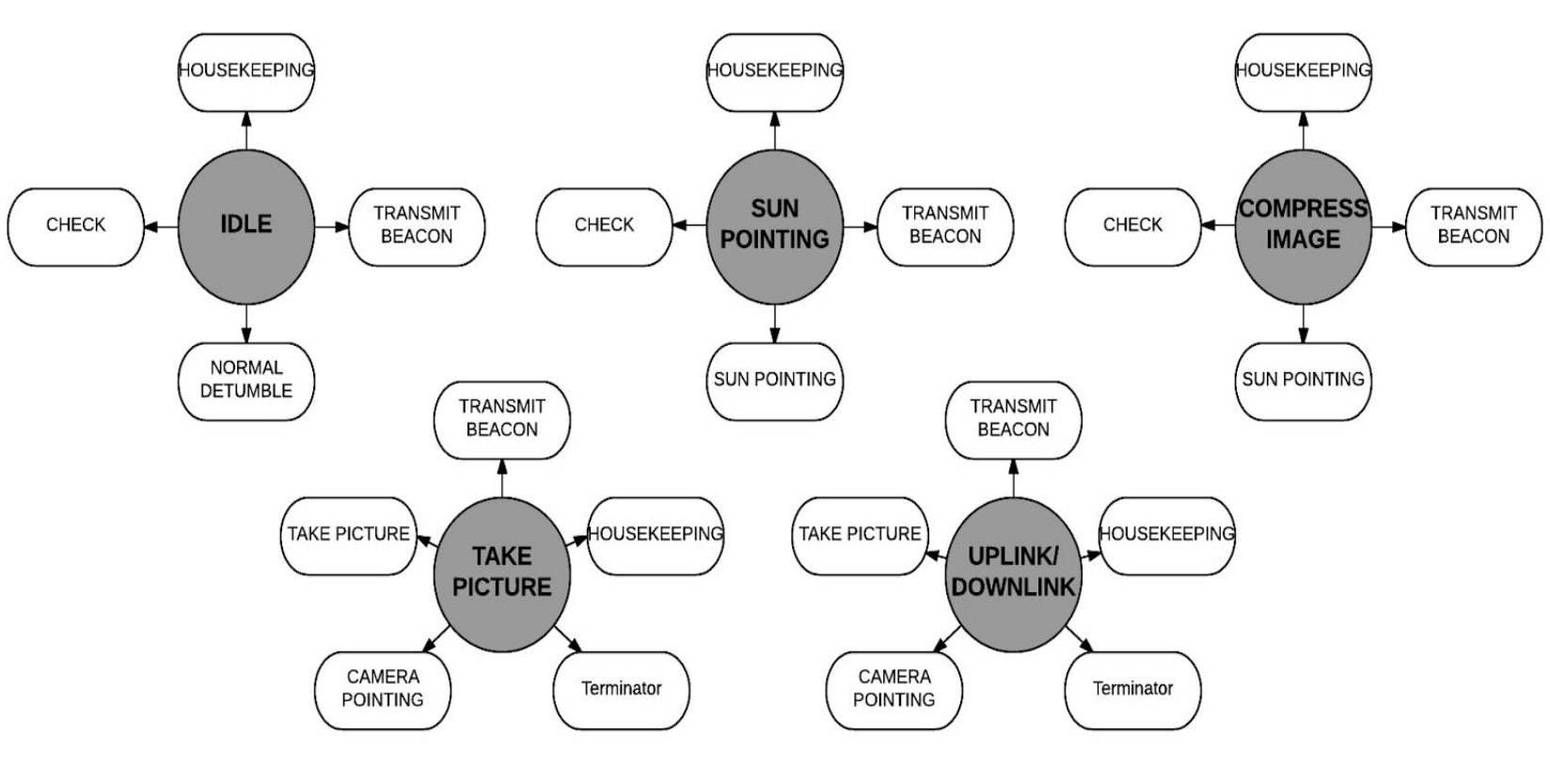

*Figure 3 : Normal Modes and associated sub-tasks*

*4.3 Userspace functions for bus access*

The lowermost layer of abstraction consists of a set of modules for device access on the specified buses. This provides the required abstraction from the device datasheet specifications while implementing the mode-wise subtasks that use the particular device. There is a set of character drivers in the kernel, some of which are developed by the team. These are accessed using the attribute functions that they export. These functions are called using preset macros within the ioctl() command. The userspace functions corresponding to the device in question implement these ioctl calls and format the data as per the usage. Such a modular design also helps incorporate fault isolation techniques.

This software design hence provides a fair amount of modularity. It also achieves the required amount of parallelism leading to an increase in efficiency. It is also possible to scale this horizontally in the middle layer for more tasks and vertically for more abstraction if need be. The flip-side is difficulty in synchronisation and mutual exclusion among the processes that form the middle layer if need arises. However, at the time of writing, an exhaustive set of tasks has been proposed that does not require mutual exclusion (other than that at kernel level which has been implemented in the device drivers or bus controller driver).

**5. Kernel modules for data transfer**

The OBC is required to gather housekeeping data from various peripherals interfaced on the I2C and SPI buses. Due to unavailability of required device drivers in the vanilla kernel, custom drivers have been developed that provide the required abstraction to the top level software layers. Drivers are represented as modules in the Linux kernel, which essentially extends the functionality of the underlying operating system [6]. The role of device drivers can be divided into two parts:

1. Communicating with the hardware to get the relevant data and storing the data in the kernel space.
2. Providing an interface between the kernel space and user space for providing the data to user applications.

The Linux kernel is organised into subsystems, each subsystem acts as a wrapper to the underlying hardware functions. A bus controller driver mediates the communication between the client drivers and actual hardware.

*5.1 User space interface*

The data can be communicated to the user space either by using the virtual filesystem, SysFs or by using the character driver API. In SysFs, every virtual file is tied to a kobject which can be used to export information from kernel space to the user space. A device driver can use SysFs to create a one-to-one correspondence between the data registers and virtual files. The user space application is hence required to do multiple file input/ output operations which can be a tedious task for an application in the absence of standard libraries such as libudev which are not present due to the usage of mdev in the kernel of the Petalinux operating system. A char driver is represented as a file in the Linux filesystem. A device can be controlled using the ioctl() calls and therefore leads to uncomplicated code in the user space. The representation of devices as files has been chosen for our purposes due to easier management and scalability of the user space code.

*5.2 SPI Controller Driver*

A flash based memory has been chosen for the storage of raw hyperspectral image data. This memory is interfaced with the OBC using the SPI bus. An SPI controller in the PL fabric acts as an interface between the memory and hyperspectral image compression IP. The SPI controller is responsible for transferring the data from flash memory to a block RAM in the FPGA. An SPI controller driver has been developed which is required to communicate to the SPI controller using the AXI interconnect between PS and PL. The driver will send the start address of the memory block in the flash memory. The SPI controller has been designed to generate the required opcodes and transfer the data in bursts. The SPI controller will generate an interrupt upon the completion of a burst data transfer. An interrupt handler divided into top and bottom halves, are tasked with sending of the next memory block address upon receiving an interrupt [7]. The top and bottom halves approach reduces the latency and effect of interrupts on the preempted task.

*5.3 Bus Protocol Drivers*

The I2C and SPI subsystems provide functionalities to the user space for communication over the I2C and SPI buses, respectively. A set of I2C client drivers were developed for the slave devices on the I2C bus [8]. Each of the slaves is represented as a node inside the bus controllers in the device tree. The SPI subsystem is divided into master and slave drivers [9]. Each device on the SPI bus is handled by a slave driver which itself uses the abstraction layer provided by the SPI master driver. The shared memory that stores the housekeeping data to be downlinked will be managed by an SPI protocol driver. This driver is linked to kernel space by the character driver interface. The memory doesn't employ a file system due to unavailability of an operating system on the telemetry micro controller unit.

## 6. Fault tolerance methods

Various methods for fault tolerance/isolation and removal are inculcated in the current model. Major points of vulnerability include, radiation tolerance of the SRAM based FPGA and flash memories, and bus failure. Furthermore, the CDH unit may also enter a non-responsive state or suffer a memory corruption. The following measures are implemented for fault tolerance as per above areas of focus. Watchdog timer shall safeguard against any and all CDH failures, and bus failures.

### *6.1 Watchdog timer*

The watchdog timer is conventionally a hardware timer that is reset ("kicked") at a predefined regular time interval. If it is not reset before it expires, the default action is to reset the hardware that was supposed to reset it. A watchdog timer exists on the Microcontroller unit of the Electrical and Power Subsystem (EPS) which is interfaced with the OBC via a GPIO pin specifically meant for this (other than the aforementioned SPI interface). If that timer is not reset by the OBC, then it is the responsibility of the EPS to perform a power cycle on the OBC. This however, is a very expensive operation. To avoid system reboots for minor faults, a hierarchy of watchdog timers has been proposed. This allows for fault isolation. The lowermost level of the hierarchy is the physical timer on the MCU of the EPS as discussed. It will be reset by sending a signal on the GPIO pin. This is critical while booting up the OBC. The timer reset for the EPS watchdog will be performed by a user space function call (which in turn calls hardware function for the GPIO interface). The software watchdog has the exclusive access to this userspace function and hence it forms the second layer of the aforementioned hierarchy. The software watchdog will be implemented as a second thread within the Flightplan process. The following reasons justify this design:

- The processes being watched are children of this process. They can easily check in at specified frequencies by sending a signal to the parent which will be handled by a specific handler built for this purpose. The handler will write to a particular memory location corresponding to each child process.
- In case the process fails to check in, the Flightplan's watchdog thread can send a SIGTERM signal to the child and terminate execution. This is possible because the linked list node contains metadata about each process which includes its process ID. SIGCHLD signal will be received by the main thread of the Flightplan as soon as child dies. It will update this information in the linked list node.

### *6.2 Radiation associated fault tolerance for FPGA*

The Single Event Effects (SEE) commonly lead to radiation related failures in the FPGA and flash memories. These are characterized into single event upset (SEU) and single event latchup (SEL). SEU occurs when a high energy radiation particle has enough energy to generate free charge carriers and change the state of a logic line and hence affect the state of a memory cell or a flip flop. SEL is said to occur when an extremely high energy radiation strikes and causes a short circuit. This leads to permanent damage to the component. As mentioned above, the current design consists of a Programmable Logic in the Zynq-7000 which is a SRAM based FPGA. SRAM-based FPGAs are sensitive to radiation found in most satellite orbits. Single Event Upsets in FPGA affect the user designed flip flops, the FPGA bit stream that is stored in the configuration memory, and some of the hidden FPGA latches, registers or the internal state [10]. The upsets that occur within the PL Fabric can be undone by resetting but the upsets encountered in the configurational memory require reconfiguration followed by a complete system reset. Some of the ways to operate FPGAs in radiation environment is to make use of active mitigation techniques and configuration scrubbing [11]. Triple-modular redundancy (TMR) is one of the active mitigating techniques which involves tripling the circuitry and inserting the voters to choose amongst the three. This method can be used for protecting against all types of single-bit failures and many multi-bit failures. Configuration scrubbing involves rewriting the configuration data periodically into the configuration memory to repair radiation-induced upsets. Though this method does not provide mitigation from radiation effects but it prevents building up of single event upsets overtime and thus breaking down of mitigation techniques like TMR. The major disadvantage of TMR is the excessive power usage and implementation complexity. The current proposal is hence to include configuration scrubbing.

*6.3 Radiation associated fault tolerance for flash memory*

Flash memories are inherently susceptible to radiation. Hence, one additional boot image is stored in an EEPROM (electrically erasable programmable read only memory). This provides for the option to update software on the flash while having a fallback in the EEPROM that has been tested heavily before launch. It is proposed to implement memory scrubbing for the flash memory storing the image data and the health metrics for downlinking. Memory scrubbing will be implemented using error correcting codes like the hamming code. This can be done by running a maintenance process in the background.